\documentstyle[11pt,epsfig]{article}
\begin{document}
\title{{\bf Late time acceleration in a deformed phase space model of dilaton cosmology}}
\author{B. Vakili$^1$\thanks{%
b-vakili@iauc.ac.ir},\,\, P. Pedram$^2$\thanks{%
pouria.pedram@gmail.com}\,\, and S. Jalalzadeh$^3$\thanks{%
s-jalalzadeh@cc.sbu.ac.ir}
\\\\
$^1${\small {\it Department of Physics, Azad University of Chalous, P.O. Box 46615-397, Chalous, Iran}} \\
$^2${\small {\it Plasma Physics Research Center, Science and Research Campus, Islamic Azad University,
Tehran, Iran }}\\
$^3${\small {\it Department of Physics, Shahid Beheshti University, C. G. Evin,
Tehran 19839, Iran}}}\maketitle

\begin{abstract}
The effects of noncommutativity on the phase space of a dilatonic
cosmological model is investigated. The existence of such
noncommutativity results in a deformed Poisson algebra between
the minisuperspace variables and their momenta conjugate. For an
exponential dilaton potential, the exact solutions in the
commutative and noncommutative cases, are presented and compared.
We use these solutions to address the late time acceleration issue
of cosmic evolution.\vspace{5mm}\newline PACS numbers: 98.80.-k,
04.20.Fy, \vspace{0.8mm}\newline Keywords: noncommutative cosmology, deformed phase space, late time acceleration\vspace{.5cm}
\end{abstract}
\section{Introduction}
Noncommutativity between space-time coordinates was first
introduced by Snyder \cite{1}, and in more recent times a great
deal of interest has been generated in this area of research
\cite{2, 4}. This interest has been gathering pace in recent years
because of strong motivations in the development of string and
M-theories. However, noncommutative theories may also be justified
in their own right because of the interesting predictions they
have made in particle physics, a few examples of which are the
IR/UV mixing and non-locality \cite{7}, Lorentz violation \cite{8}
and new physics at very short distance scales \cite{8}-\cite{10}.
Noncommutative versions of ordinary quantum \cite{11} and
classical mechanics {\cite{12, 13}} have also been studied and
shown to be equivalent to their commutative versions if an
external magnetic field is added to the Hamiltonian. A different
approach to noncommutativity is through the introduction of
noncommutative fields \cite{14}, that is, fields or their
conjugate momenta are taken as noncommuting. These effective
theories can address some of the problems in ordinary quantum
field theory, e.g. regularization \cite{14} and predict new
phenomenon, such as Lorentz violation \cite{15}, considered as one
of the general predictions of quantum gravity theories \cite{16}.

In cosmological systems, since the scale factors, matter fields
and their conjugate momenta play the role of dynamical variables
of the system, introduction of noncommutativity by adopting the
approach discussed above is particularly relevant. The resulting
noncommutative classical and quantum cosmology of such models have
been studied in different works \cite{17}. These and similar works
have opened a new window through which some of problems related to
cosmology can be looked at and, hopefully, resolved. For example,
an investigation of the cosmological constant problem can be found
in \cite{18}. In \cite{19} the same problem is carried over to the
Kaluza-Klein cosmology. The problem of compactification and
stabilization of the extra dimensions in multidimensional
cosmology may also be addressed using noncommutative ideas in
\cite{20}.

In recent years many efforts have been made in cosmology from
string theory point of view \cite{21}-\cite{24}. In the pre-big
bang scenario, based on the string effective action, the birth of
the universe is described by a transition from the string
perturbative vacuum with weak coupling, low curvature and cold
state to the standard radiation dominated regime, passing through
a high curvature and strong coupling phase. This transition is
made by the kinetic energy term of the dilaton, a scalar field
with which the Einstein- Hilbert action of general relativity is
augmented, see \cite{26} for a more modern review.

In this Letter we deal with noncommutativity in a dilaton
cosmological model with an exponential dilaton potential and to
facilitate solutions for the case under consideration, we choose a
suitable metric. Our approach to noncommutativity is through its
introduction in phase space constructed by minisuperspace fields
and their conjugate momenta. Indeed, in general relativity
formulation of gravity in a noncommutative geometry of space-time
is highly nonlinear and setting up cosmological models is not an
easy task. Here our aim is to study some aspects regarding the
application of noncommutativity in cosmology, i.e. in the context
of a minisuperspace reduction of the dynamics.  Since our model
has two degrees of freedom, the scale factor $a$ and the dilaton
$\phi$, with a change of variables, we have a set of dynamical
variables $(u,v)$, which are suitable candidates for introducing
noncommutativity in the phase space of the problem at hand. We
present exact solutions of commutative and noncommutative
cosmology and show that commutative model cannot describe the late
time acceleration while the noncommutative counterpart of the
model clearly points to a possible late time acceleration.
\section{The Model}
In $D=4$ dimension lowest order gravi-dilaton effective action, in
the string frame, can be written as \cite{28}
\begin{equation}\label{A}
{\cal S}=-\frac{1}{2 \lambda_s^2}\int d^4x
\sqrt{-g}e^{-\phi}\left({\cal R}+\partial_{\mu}\phi
\partial^{\mu}\phi+V(\phi)\right),\end{equation}where $\phi$ is
the dilaton field, $\lambda_s$ is the fundamental length parameter of string theory and $V(\phi)$ is the dilaton potential. In the
string frame our fundamental unit is the string length $l_s$, and
thus the Planck mass, which is the effective coefficient of the
Ricci scalar ${\cal R}$, varies with the dilaton. One can also
write the action in the Einstein frame, for which the fundamental
unit is the Planck length. Since the Planck length is more
appropriate for our purpose, we prefer to work in the Einstein
frame. In \cite{24}, it is shown in detail that action (\ref{A})
in the Einstein frame takes the form
\begin{equation}\label{B}
{\cal S}=-\frac{M_4^2}{2}\int d^4x \sqrt{-g}\left({\cal
R}-\frac{1}{2}\partial_{\mu}\phi
\partial^{\mu}\phi-V(\phi)\right),\end{equation}where now all
quantities in the action are in the Einstein frame. We consider a
spatially flat FRW space-time which, following \cite{29}, is
specified by the metric
\begin{equation}\label{C}
ds^2=-\frac{N^2(t)}{a^2(t)}dt^2+a^2(t)\delta_{ij}dx^idx^j.\end{equation}Here
$N(t)$ is the lapse function and $a(t)$ represents the scale
factor of the universe. The square of the scale factor dividing
the lapse function turns out to simplify the calculations and
makes the Hamiltonian quadratic. Now, it is easy to show that the
effective Lagrangian of the model can be written in the minisuperspace
$Q^A=(a,\phi)$ in the form
\begin{equation}\label{D}
{\cal L}=\frac{1}{N}\left(-\frac{1}{2} a^2
\dot{a}^2+\frac{1}{2}a^4
\dot{\phi}^2\right)-Na^2V(\phi).\end{equation}
The momenta conjugate to the dynamical variables are given by
\begin{equation}\label{D1}
p_a=\frac{\partial {\cal L}}{\partial \dot{a}}=-\frac{1}{N}a^2\dot{a},\hspace{0.5cm}
p_{\phi}=\frac{\partial {\cal L}}{\partial \dot{\phi}}=\frac{1}{N}a^4\dot{\phi},\end{equation}leading to the following Hamiltonian
\begin{equation}\label{D2}
H=\frac{1}{2}{\cal G}^{AB}P_AP_B+{\cal U}(Q^A)=\frac{N}{2a^2}\left(-p_a^2+\frac{1}{a^2}p_{\phi}^2\right)+Na^2V(\phi).\end{equation}
Now, it is easy to see that the corresponding minisuperspace has the following minisuper
metric
\begin{equation}\label{D3}
{\cal
G}_{AB}dQ^AdQ^B=\frac{a^2}{N}\left(-da^2+a^2d\phi^2\right).\end{equation}
To apply the deformed commutators to the dynamical variables in a
minisuperspace which is represented by a curved manifold with a
minisuper metric given by (\ref{D3}), in a natural way, one has to
deal with some difficulties which the most important of them is
the ambiguity in the ordering of factors $Q$ and $P_Q$. Therefore,
the above minisuperspace does not have the desired form for
introducing noncommutativity among its coordinates. To avoid the
physical difficulties and simplify the model, consider the
following change of variables $Q^A =(a,\phi)\rightarrow
q^A=(u,v)$, \cite{30}
\begin{equation}\label{E}
u=\frac{a^2}{2}\cosh \alpha \phi,\hspace{.5cm}v=\frac{a^2}{2}\sinh
\alpha \phi,\end{equation}where $\alpha$ is a positive constant.
In terms of these new variables the Lagrangian (\ref{D}) takes the
form
\begin{equation}\label{F}
{\cal
L}=\frac{1}{2N}\left(\dot{v}^2-\dot{u}^2\right)-2N\left(u-v\right)e^{\alpha
\phi}V(\phi).\end{equation}From now on, we choose an exponential
potential
\begin{equation}\label{G}
V(\phi)=\frac{V_0}{2}e^{-\alpha \phi},\end{equation}which
simplifies the last term in the Lagrangian (\ref{F}) leading to
\begin{equation}\label{H}
{\cal
L}=\frac{1}{2N}\left(\dot{v}^2-\dot{u}^2\right)-NV_0\left(u-v\right),\end{equation}with the corresponding Hamiltonian becoming
\begin{equation}\label{I}
H=N{\cal
H}=N\left[-\frac{1}{2}p_u^2+\frac{1}{2}p_v^2+V_0\left(u-v\right)\right].\end{equation}Thus,
in the minisuperspace constructed by $q^A=(u,v)$, the metric, for a constant $N$, is Minkowskian and represented by
\begin{equation}\label{I1}
\bar{{\cal
G}}_{AB}dq^Adq^B=N\left(-du^2+dv^2\right).\end{equation}Now, we
have a set of variables $(u,v)$ endowing the minisuperspace with a
Minkowskian metric and hence this set of dynamical variables are
suitable candidates for introducing noncommutativity in the phase
space of the problem at hand. The preliminary setup for describing
the model is now complete. In what follows, we will study the
classical cosmology of the minisuperspace model described by
Hamiltonian (\ref{I}) in commutative and noncommutative
frameworks.
\section{Cosmological dynamics}
The classical solutions of the model described by Hamiltonian
(\ref{I}) can be easily obtained. Since our aim here is to compare
the commutative solutions with their noncommutative counterparts,
in what follows we consider commutative and noncommutative
classical cosmologies, and compare the results with each other.
\subsection{Commutative case}
As is well known for a dynamical system with phase space variables
$(x_i, p_i)$, the Poisson algebra is described by the following
Poisson brackets
\begin{equation}\label{J}
\left\{x_i,x_j\right\}=\left\{p_i,p_j\right\}=0,\hspace{.5cm}\left\{x_i,p_j\right\}=\delta_{ij},\end{equation}where
$x_i(i=1,2)=u,v$ and $p_i(i=1,2)=p_u, p_v$. Therefore, the
equations of motion become (in $N=1$ gauge)
\begin{equation}\label{K}
\dot{u}=\left\{u,N{\cal
H}\right\}=-p_u,\hspace{.5cm}\dot{p_u}=\left\{p_u,N{\cal
H}\right\}=-V_0,\end{equation}
\begin{equation}\label{L}
\dot{v}=\left\{v,N{\cal
H}\right\}=p_v,\hspace{.5cm}\dot{p_v}=\left\{p_v,N{\cal
H}\right\}=V_0,\end{equation} Equations (\ref{K}) and (\ref{L})
can be immediately integrated to yield
\begin{equation}\label{M}
u(t)=\frac{1}{2}V_0
t^2-p_{0u}t+u_0,\hspace{.5cm}p_u(t)=-V_0t+p_{0u},\end{equation}
\begin{equation}\label{N}
v(t)=\frac{1}{2}V_0
t^2+p_{0v}t+v_0,\hspace{.5cm}p_v(t)=V_0t+p_{0v}.\end{equation}
Now, these solutions must satisfy the zero energy condition,
${\cal H}=0$. Thus, substitution of equations (\ref{M}) and
(\ref{N}) into (\ref{I}) gives a relation between integration
constants as
\begin{equation}\label{O}
p_{0v}^2-p_{0u}^2=2V_0 (v_0-u_0).\end{equation}Equations (\ref{M})
and (\ref{N}) are like the equation of motion for a particle
moving in a plane with its acceleration components equal to $V_0$,
while $-p_u(t)$ and $p_v(t)$ play the role of its velocity components.
Finally, using equation (\ref{E}), the scale factor and dilaton
take the following forms
\begin{equation}\label{P}
a(t)=\sqrt{2}\left[-V_0(p_{0u}+p_{0v})t^3+3V_0(u_0-v_0)t^2-2(u_0p_{0u}+v_0p_{0v})t+(u_0^2-v_0^2)\right]^{1/4},\end{equation}
\begin{equation}\label{R}
\phi(t)=\frac{1}{2\alpha}\ln \frac{V_0t^2+(p_{0v}-p_{0u})t+(u_0+v_0)}{-(p_{0u}+p_{0v})t+(u_0-v_0)}.\end{equation}
The limiting behavior of $a(t)$ and $\phi(t)$ in the early and late
times is then as follows
\begin{eqnarray}\label{S}
\left\{
\begin{array}{ll}
a(t)\sim t^{1/4},\hspace{.5cm}\phi(t)\sim
\mbox{const.},\hspace{.5cm}t\ll1,\\\\
a(t)\sim t^{3/4},\hspace{.5cm}\phi(t)\sim \ln t,\hspace{1cm}t\gg1.
\end{array}
\right.
\end{eqnarray}
We see that in the usual commutative phase space of our model the
scale factor has a decelerated expansion in early times while it also
undergoes a decelerated phase in its late time evolution due to a
constant and growing with time dilatonic field, respectively. The
evolution of the universe based on (\ref{S}) begins with a
big-bang singularity at $t=0$ and follows the power law expansion
$a(t)\sim t^{3/4}$ at late time of cosmic evolution (which is not consistent with
late time observations) while the dilaton
has a monotonically increasing behavior coming from a constant
value and logarithmically blows up at late time. We shall see in the next subsection how this classical picture will be modified by introducing the deformed phase space model.

\subsection{Deformed phase space model}
An important ingredient in any model theory related to the quantization of a cosmological setting is the choice of the quantization
procedure used to quantize the system. The most widely used method has traditionally been
the canonical quantization method based on the Wheeler-DeWitt equation, which is
nothing but the application of the Hamiltonian constraint to the wavefunction of the universe.
A particularly interesting but rarely used approach to study the quantum effects is to introduce
a deformation in the phase space of the system. It is believed that such a deformation of
phase space is an equivalent path to quantization, in par with other methods, namely canonical
and path integral quantization \cite{31}. This method is based on the Wigner quasi-distribution
function and Weyl correspondence between quantum mechanical operators in Hilbert space
and ordinary c-number functions in phase space. The deformation in the usual phase space
structure is introduced by Moyal brackets which are based on the Moyal product.
However, to introduce such deformations it is more convenient to work with Poisson brackets
rather than Moyal brackets. From a cosmological point of view, models are built in a minisuperspace. It is therefore
safe to say that studying such a space in the presence of deformations mentioned above can be
interpreted as studying the quantum effects on cosmological solutions.

Let us now proceed to study the behavior of the above model in a
deformed phase space framework such that the minisuperspace
variables do not (Poisson) commute with each other. In general,
noncommutativity between phase space variables can be understood
by replacing the usual product with the star product, also known
as the Moyal product law between two arbitrary functions of
position and momentum, as
\begin{equation}\label{35}
(f\star_\alpha
g)(x)=\exp\left[\frac{1}{2}\alpha^{ab}\partial_a^{(1)}\partial_b^{(2)}\right]
f(x_1)g(x_2)|_{x_1=x_2=x},
\end{equation}
such that
\begin{eqnarray}\label{36}
\alpha_{ab}=\left(%
\begin{array}{cc}
\theta_{ij}  & \delta_{ij}+\sigma_{ij} \\
-\delta_{ij}-\sigma_{ij} & \beta_{ij}\\
\end{array}%
\right),
\end{eqnarray}
where the $2\times 2$ matrices $\theta$ and $\beta$ are assumed to
be antisymmetric and represent the noncommutativity in coordinates and
momenta, respectively. Also, $\sigma$ can be written as a combination of $\theta$ and
$\beta$. With this product law, the $\alpha$-deformed Poisson brackets can be
written as
\begin{equation}\label{37}
\{f,g\}_\alpha=f\star_\alpha g-g\star_\alpha f.
\end{equation}
Upon a simple calculation we have
\begin{equation}\label{38}
\{x_i,x_j\}_\alpha=\theta_{ij}, \hspace{0.5cm}
\{x_i,p_j\}_\alpha=\delta_{ij}+\sigma_{ij}, \hspace{0.5cm}
\{p_i,p_j\}_\alpha=\beta_{ij} \hspace{.15cm}.
\end{equation}
It is worth noting at this stage that in addition to
noncommutativity in position variables, we have also considered
noncommutativity in the corresponding momenta. Such a noncommutativity can be motivated by string theory corrections to the
Einstein gravity. This should be
interesting since its existence is in fact due essentially to the
existence of noncommutativity on the space sector
and it would somehow be natural to include it in our considerations. In this work we consider a deformed phase space in
which $\theta_{ij}=-\theta \epsilon_{ij}$ and $\beta_{ij}=-\beta \epsilon_{ij}$, where
 $\epsilon_{ij}$ are the Levi-Civita symbols.

Now, consider the following transformation on the classical phase
space $\{u,v,p_u,p_v\}$
\begin{eqnarray}\label{T4}
\left\{
\begin{array}{ll}
\hat{v}=v-\frac{\theta}{2}p_u, \hspace{0.5cm}
\hat{u}=u+\frac{\theta}{2}p_v,\\\\
\hat{p}_v=p_v+\frac{\beta}{2}u, \hspace{0.5cm}
\hat{p}_u=p_u-\frac{\beta}{2}v.
\end{array}
\right.
\end{eqnarray}
It can easily be checked that if $\{u,v,p_u,p_v\}$ obey the usual
Poisson algebra (\ref{J}), then
\begin{eqnarray}\label{40}
\{\hat{v},\hat{u}\}_{P}=\theta,\hspace{0.5cm}
\{\hat{u},\hat{p}_u\}_{P}=\{\hat{v},\hat{p}_v\}_{P}=1+\sigma,\hspace{0.5cm}
\{\hat{p}_v,\hat{p}_u\}_{P}=\beta,
\end{eqnarray}
where $\sigma=\frac{1}{4}\theta \beta$. These commutation relations
are the same as (\ref{38}). Consequently, for introducing
noncommutativity, it is more convenient to work with Poisson
brackets (\ref{40}) than $\alpha$-star deformed Poisson brackets
(\ref{38}). It is important to note that the relations represented
by equations (\ref{38}) are defined in the spirit of the Moyal
product given above. However, in the relations defined in
(\ref{40}), the variables $\{u,v,p_u,p_v\}$ obey the usual Poisson
bracket relations so that the two sets of ordinary and deformed
Poisson brackets represented by relations (\ref{38}) and (\ref{40})
should be considered as distinct.

With the noncommutative phase space defined above, we consider the
Hamiltonian of the noncommutative model as having the same
functional form as equation (\ref{I}), but in which the dynamical
variables satisfy the above deformed Poisson brackets, that is
\begin{equation}\label{new-hamil}
\hat{H}=N\hat{{\cal H}}=N\left[{1\over 2} (-\hat{p}_u^2 +\hat{p}_v^2) +V_0
(\hat{u}-\hat{v})\right].
\end{equation}
In physical point of view this is because that the Hamiltonian
constraint is the result of time re-parametrization invariance
which remains even when the noncommutativity is turned on. Using
the transformations (\ref{T4}), we can write the Hamiltonian
without the hat variables, that is, those that satisfy the usual
commutation relations, as
\begin{equation}\label{hat-hamil}
H=N{\cal H}=N\left[\frac{1}{2}(p_v^2-p_u^2)+\frac{1}{8}
\beta^2(u^2-v^2)+\frac{1}{2}\beta(up_v+vp_u)+\frac{V_0}{2}\theta(p_v+p_u)+V_0(u-v)\right].
\end{equation}
The equations of motion (again in $N=1$ gauge) can now be written easily with respect to
Hamiltonian (\ref{hat-hamil})
\begin{eqnarray}\label{O}
\left\{
\begin{array}{ll}
\dot{u}=\{u,H\}_P=-p_u+\frac{1}{2}\beta v+\frac{V_0}{2}\theta,\\\\
\dot{v}=\{v,H\}_P=p_v+\frac{1}{2}\beta u+\frac{V_0}{2}\theta,\\\\
\dot{p}_u=\{p_u,H\}_P=-\frac{1}{4}\beta^2 u-\frac{1}{2}\beta p_v-V_0,\\\\
\dot{p}_v=\{p_v,H\}_P=\frac{1}{4}\beta^2 v-\frac{1}{2}\beta p_u+V_0,
\end{array}
\right.
\end{eqnarray}
Eliminating the momenta from the above equations, we get
\begin{eqnarray}
\left\{
\begin{array}{ll}
\ddot{v}-\beta \dot{u}-V_0(1-\sigma)=0, \\\\
\ddot{u}-\beta \dot{v}-V_0(1-\sigma)=0.\end{array}
\right.
\end{eqnarray}We see that the deformation parameters appear as a coupling constant between equations
of motion for $u$ and $v$. Let us define the new variables $\eta=v+u$ and $\zeta=v-u$, in terms of which we have

\begin{eqnarray}
\left\{
\begin{array}{ll}
\ddot{\eta}-\beta \dot{\eta}-2V_0(1-\sigma)=0, \\\\
\ddot{\zeta}+\beta \dot{\zeta}=0.\end{array}
\right.
\end{eqnarray}
Integrating these equations yields
\begin{eqnarray}
\left\{
\begin{array}{ll}
\eta(t)=A' e^{\beta t}-\frac{2V_0}{\beta}(1-\sigma)t+C',\\\\
\zeta(t)=B' e^{-\beta t}+D',\end{array}
\right.
\end{eqnarray}
where $A', B', C', D'$ are integration constants. Going back to the variables $v$ and $u$, we obtain
\begin{eqnarray}
\left\{
\begin{array}{ll}
v(t)=A e^{\beta t}+B e^{-\beta t}-\frac{V_0}{\beta}(1-\sigma)t+C,\\\\
u(t)=A e^{\beta t}-B e^{-\beta t}-\frac{V_0}{\beta}(1-\sigma)t+D.\end{array}
\right.
\end{eqnarray}
Here, $A, B, C, D$ are linear combinations of the prime constants.
The requirement that the deformed Hamiltonian constraints
(\ref{hat-hamil}) should hold during the evolution of the system
leads to the following relation between these integrating
constants

\begin{equation}
2\beta \,A\,B+\frac{V_0}{\beta}(1-\sigma)(C-D)=0.
\end{equation}
To proceed further, we can choose a particular set of constants, say $B=-A$. Now, $v(t)$, $u(t)$, and Hamiltonian constraint take
the form
\begin{eqnarray}
\left\{
\begin{array}{ll}
v(t)=2A \sinh(\beta t)-\frac{V_0}{\beta}(1-\sigma)t+C,\\\\
u(t)=2A \cosh(\beta t)-\frac{V_0}{\beta}(1-\sigma)t+D,\\\\
2\beta \,A^2=\frac{V_0}{\beta}(1-\sigma)(C-D).\end{array}
\right.
\end{eqnarray} Now, let us return to the variables $a(t)$ and
$\phi(t)$ using the transformation (\ref{E}), in terms of which we obtain the
corresponding deformed classical cosmology as
\begin{eqnarray}
\left\{
\begin{array}{ll}
a(t)=2\left[(1+\beta
t)A^2+\left(D-\frac{V_0}{\beta}(1-\sigma)t\right)A\cosh\beta
t-\left(C-\frac{V_0}{\beta}(1-\sigma)t\right)A\sinh\beta t+\frac{1}{4}(D^2-C^2)\right]^{\frac{1}{4}},\\\\
\phi(t)=\frac{1}{2\alpha}\ln\left(\frac{2Ae^{\beta
\,t}-\frac{2V_0}{\beta}(1-\sigma)t+D+C}{2Ae^{-\beta
\,t}+D-C}\right).
\end{array}
\right.
\end{eqnarray}
As in the previous section, we may express the early and late time limiting behavior of the scale factor. The scale factor $a$ in these
two regimes takes the form
\begin{eqnarray}\label{U}
\left\{
\begin{array}{ll}
a(t)\sim \left[\frac{V_0}{\beta}(1-\sigma)t\right]^{1/4},\hspace{0.5cm}t\ll1,\\\\
a(t)\sim e^{\frac{\beta
t}{4}},\hspace{2.6cm}t\gg1.
\end{array} \right.
\end{eqnarray}
This means that the scale factor begins again from a big-bang singularity and then behaves as the usual commutative case (\ref{S}) in the early times. However, the differences between the commutative and noncommutative cases are manifest in the late time behavior where in the noncommutative case with $t\rightarrow \infty$ the scale factor becomes an exponentially function, for which (\ref{C}) represents a de-Sitter metric with a cosmological constant $\Lambda=\frac{3}{16}\beta^2$. Therefore, for late times the behavior of the scale factor becomes exponential, pointing to the existence of a deformation parameter in the corresponding phase space of cosmological setting which can be interpreted as a candidate for dark energy. Taking the relation for $\Lambda$ given above, the value of the deformation parameter $\beta$ can be estimated. If the cosmological constant is taken as $\Lambda \sim 10^{-52}m^{-2}$, a generally accepted value determined by observation, the deformation parameter is found to be $\beta \sim 10^{-26} m^{-1}$, which is comparable with the upper bounds for such noncommutative parameters resulting from theoretical predictions and experimental data obtained
in the field theory and quantum gravity \cite{32}.

\begin{figure}
\begin{tabular}{c}\epsfig{figure=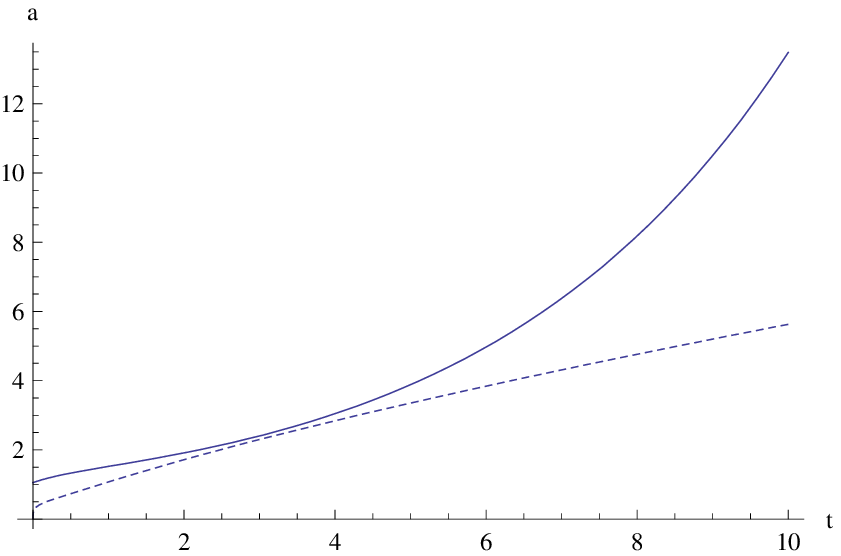,width=7cm}
\hspace{2cm} \epsfig{figure=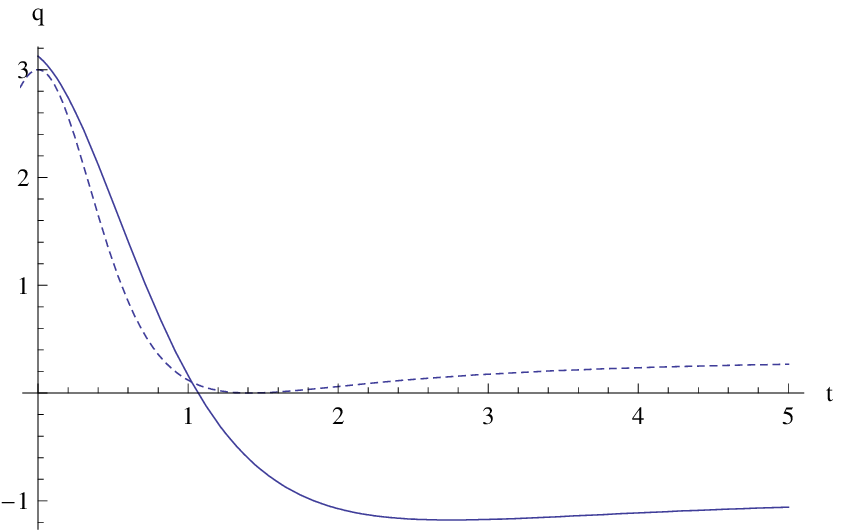,width=7cm}
\end{tabular}
\caption{\footnotesize  Qualitative behavior of the scale factor (left) and the deceleration parameter (right) as a
function of time. The dashed lines represent these quantities for the usual commutative case while the solid lines represent their behavior in the
deformed phase space cosmology. The figures are plotted for numerical
values $A=1$, $\beta=1$, $C=-2$ and $D=1$. After examining some
other values for these parameters, we verify that the general
behavior of the curves is repeated.}\label{fig1}
\end{figure}

We can also compute the deceleration parameter $q=-\frac{a\ddot{a}}{\dot{a}^2}$ for these two models. As is well known the deceleration parameter indicates by how much the expansion of the universe is slowing down.
If the expansion is speeding up, for which there appears to be
some recent evidence, then this parameter will be negative. In
figure \ref{fig1}, we have shown the approximate behavior of the scale factor $a(t)$ and the deceleration parameter
$q(t)$ for typical values of the parameters in both commutative and noncommutative regimes. It is seen that the behavior of two scale factors is the same in the early times and the minor differences may be interpreted by the fact that the early time behavior of the commutative scale factor will be determined only by the initial conditions, see (\ref{P}), while for the noncommutative one, the quantum effects are presence as well, see the first relation of (\ref{U}). On the other hand, as is clear from the figure, in contrast to the commutative case for which the deceleration parameter is positive in all times, for the deformed phase space case we have negative acceleration (positive deceleration parameter) at early times so that the universe decelerates its expansion in this era, and positive acceleration (negative deceleration parameter) for late times
which means that the universe currently accelerates its expansion.
Therefore, the late time acceleration, suggested by recent
supernova observation, can be addressed by considering a deformed
phase space in the corresponding cosmological set up.
\section{Conclusions}
In this letter we have studied the effects of noncommutativity in phase space,
on classical cosmology of a dilaton model with an exponential dilaton potential. Motivation of such study is that the construction of an effective theory is usually the result of the observation that the corresponding
full theory is too complicated to deal with. This, for example, is true in describing the
quantum effects on cosmology since the full theory is immensely difficult to handle. Introduction of a deformed phase space
can be interpreted as one such effective theory. In the present study, in the case of commutative
phase space, the evolution of the classical universe is like the motion of a particle (universe) moving
on a plane with a constant acceleration. We saw that in this case the universe decelerates its expansion both in early and late times of cosmic evolution, in contrast to the current observation data. We have investigated the possibility of having a late
time accelerated phase of the universe, suggested by recent
supernova observation, in the context of a deformed phase space
model of string cosmology action. Indeed, we have presented a
scenario in which cosmic acceleration occurs late in the history
of the universe due to introduction of noncommutativity in phase
space, on classical cosmology of a dilaton model with an
exponential dilaton potential. We have found that while the usual
classical model cannot support this acceleration, a classical
model with noncommutative phase space variable can drive this late
time acceleration.
\vspace{5mm}\newline \noindent {\bf
Acknowledgement}\vspace{2mm}\noindent\newline
B. Vakili is grateful to the research council of Azad University of Chalous for
financial support.

\end{document}